\documentclass[sigconf]{acmart}

\copyrightyear{2026}
\acmYear{2026}
\setcopyright{acmcopyright}\acmConference[ApPLIED '26]{Proceedings of the 2026
Workshop on Advanced tools, programming languages, and PLatforms for
Implementing and Evaluating algorithms for Distributed systems}{July 6, 2026}
{Egham, England}
\acmBooktitle{Proceedings of the 2026 Workshop on Advanced tools, programming
languages, and PLatforms for Implementing and Evaluating algorithms for
Distributed systems (ApPLIED '26), July 6, 2026, Egham, England}
\acmPrice{15.00}
\acmDOI{10.1145/3524053.3542744}
\acmISBN{978-1-4503-9280-8/22/07}

\usepackage[utf8]{inputenc}
\usepackage{comment}
\usepackage{listings}
\usepackage{xcolor}
\usepackage{enumitem,kantlipsum}
\usepackage{booktabs}
\usepackage{pgfplots}
\usepgfplotslibrary{groupplots}
\pgfplotsset{compat=1.18}

\title{QUANTAS~2\\ An Abstract, Concrete and Byzantine Simulator}
\author{Joseph~Oglio} 
\email{joglio@rio.edu}
\affiliation{\institution{Rio Grande University} \country{Rio Grande, USA}}
\author{Mikhail~Nesterenko}
\email{mikhail@cs.kent.edu}
\affiliation{\institution{Kent State University} \country{Kent, USA}}
\begin{document}


\begin{abstract} We present QUANTAS~2: a new distributed algorithm simulator and quantitative  performance analysis tool.  
We use the original QUANTAS as a foundation. QUANTAS~2 can perform fast abstract exploration, concrete validation, and adversarial fault injection while preserving a compact implementation model for distributed algorithm researchers.

The original QUANTAS was designed as an abstract, round-based simulator, which allows researchers to separate algorithmic behavior from the artifacts of a particular operating system, network stack, or physical deployment.  
QUANTAS~2 extends that design in two directions. First, QUANTAS~2 supports a concrete socket-based execution mode, allowing the same algorithm implementations and JSON experiment descriptions to run across local or distributed computers. Second, QUANTAS~2 adds a reusable Byzantine-fault interface in which Byzantine behavior is encoded as composable fault strategy that substitutes correct sends, receives, and local computation. This allows researchers to simulate crash, equivocation, selfish-mining, and other adversarial behaviors without rewriting the simulated algorithm.

We demonstrate the resulting platform on blockchain, consensus, distributed hash table, and reliable data link algorithms. We perform parasite-chain sweeps for proof-of-work blockchains, PBFT equivocation experiments, Raft crash experiments, and Chord/Kademlia scale experiments over both abstract and concrete modes. 
\end{abstract}
\maketitle

\section{Introduction}

Distributed algorithm simulation is essential for exploring its properties and behavior~\cite{adamek17jpdc,chord,zamani2018rapidchain,PBFT}. % Joseph, cite simulations of algorithms
QUANTAS~\cite{oglio2022quantitative,QUANTASgithub} is a popular simulator used in a number of research efforts~\cite{blockguard,hood2021partitionable,bricker2022blockchain,oglio2024trail,oglio2025smartshards,hicks2025can,bricker2024consensus,oglio2025torus}. This simulator is fast, flexible, easy to learn and relatively scalable. However, QUANTAS limits algorithm simulation to the abstract model. That is, distributed processes and their communication are simulated through a sequence of computation rounds on a single computer. 

Abstract simulation allows researchers to focus purely on the algorithm's behavior and simulate arbitrary communication and computation patterns. However, the practical aspects of system operation and interaction with the environment are not simulated. Therefore, it is unclear how the algorithm performs in reality. To study the behavior of a complete distributed system, researchers are forced to re-implement their algorithm in the concrete environment.

This effectively requires two separate efforts. Furthermore, it is uncertain if the abstract and the concrete algorithm implementations are equivalent. Moreover, if the researchers compare their new algorithm with existing algorithms, they have to implement all of them for their simulation. This, raises questions of the veracity of their particular implementation of the original algorithms. 
% cite some papers that have concrete simulation that don't have abstract: 
% Checked
%Rapdichain, Alrogrand?, Chord? Kademlia?

Distributed algorithms are designed to tolerate system faults and adversarial attacks. A Byzantine process~\cite{lamport2019byzantine,pease1980reaching} is allowed to behave arbitrarily. This is the most general process fault model. It is therefore frequently considered. However, simulating such faults is challenging. Indeed, just making a process execute random actions is relatively easy to counteract. Yet, Byzantine processes tend to be at their most destructive when they launch a coordinated attack that appears realistic to the other processes. Simulation of this kind of complex adversarial behavior is seldom done.

% find sebastien's paper on byzantine fault injection or some Sebastien's papers

In this paper, we present QUANTAS~2, a simulator that uses the original QUANTAS as a foundation and extends it to provide both concrete and abstract implementation of the same algorithm as well as Byzantine fault injection. Specifically, in the abstract mode, QUANTAS~2 uses simulated channels for message transmission. In concrete mode, the same processes may use local or distributed sockets to communicate to other simulated processes.
%
%This is not intended to replace physical testbeds or full network emulators.
QUANTAS~2 thus helps researchers to bridge the gap between abstract algorithm simulation and deployment-oriented evaluation. It answers a practical question of whether the abstract algorithm specification can also execute as a real distributed program with the same algorithm objects, message encodings, and experiment parameters.

%This is not intended to replace physical testbeds or full network emulators.

QUANTAS~2 has a Byzantine functionality layer: a user can attach explicit fault strategies to selected peers, including crash, equivocation, message rewriting, message withholding, and local-computation replacement. QUANTAS~2 has improved structured logging and experiment output, so that large parameter sweeps can be reproduced and post-processed without manually scraping console traces. 
% added this, MN
The experiment setup is shared by concrete and abstract level. This makes it easy to conduct identical experiments in both modes.
These features make QUANTAS~2 useful not only for measuring benign executions, but also for investigating robust algorithms such as Byzantine fault-tolerant consensus and proof-of-work blockchain attacks.

\ \\ 
The remainder of this paper is organized as follows.
We detail related work. We, then, describe the dual abstract/concrete architecture of QUANTAS~2 and show how experiment configurations are shared across both modes. We introduce a reusable fault-injection interface for Byzantine behavior and explain how faults intercept sends, receives, and computation. We evaluate representative benign and faulty executions, including parasite-chain proof-of-work sweeps, PBFT equivocation, Raft crash tolerance, and sparse-topology DHT scale experiments in both execution modes.

% this is  related work, MN
%We position QUANTAS~2 with respect to distributed-system simulators, blockchain simulators, and recent BFT testing approaches such as Twins~\cite{bano2020twins}.
\section{Related Work}

QUANTAS~2 is related to three lines of work. First, it is related to network simulators and emulators such as ns-3~\cite{ns3}, OMNET++~\cite{omnetpp}, and Mininet~\cite{lantz2010network}. These tools are powerful when packet-level, protocol-stack, or software-defined-network behavior is the object of study. Onlike classic network simulators, QUANTAS~2 abstracts low-level network details so that distributed-algorithm researchers can configure salient algorithmic parameters, topologies, and faults with greater simplicity.

Second, QUANTAS~2 is related to peer-to-peer and overlay simulators such as PeerSim~\cite{peersim}, Sinalgo~\cite{sinalgo}, and JBotSim~\cite{jbotsim}. These systems are useful for large distributed algorithms and education. Instead, QUANTAS~2 contributes a compact C++ implementation model, JSON experiment descriptions, concrete socket execution, and explicit fault strategies for adversarial behavior.

Third, QUANTAS~2 is related to BFT testing and fault-injection work. BFTSim studied Byzantine algorithms under configurable network and computational costs~\cite{singh2008bft}. Twins~\cite{bano2020twins} generates Byzantine scenarios by duplicating nodes with the same identity. ByzzFuzz~\cite{winter2023randomized} and related approaches use randomized or systematic Byzantine fault injection to find bugs. QUANTAS~2 is complementary to these systems. Our simulator's goal is not only algorithm debigging, but quantitative experimentation: once a fault model is encoded, the same model can be applied to various network sizes, topologies, delay distributions, quorum sizes, mining powers,  recovery schedules and other experiment parameters.

\section{Simulator Design Principles,\\ Architecture and Setup}

\subsection{Design Principles and Terms}

\textbf{Dual execution.} Abstract simulation is fast and controllable, while concrete execution exposes the implementation to process scheduling, sockets, deployment scripts, and host-level failures. QUANTAS~2 supports both modes. The user selects the mode at build or run-time. The same peer code and experiment configuration can then be used for abstract rounds or concrete socket communication. This keeps the fast compile-test-analysis cycle of abstract simulation while creating a path toward implementation validation.

\ \\
\textbf{Fault injection.} Distributed algorithms are often specified under crash, omission, timing, or Byzantine fault assumptions. In QUANTAS~2 faults are implemented as reusable objects that can intercept messages and computation at the peer boundary. This design lets the researcher vary the adversary independently from the algorithm and topology.

\ \\
\textbf{Observability.} Reproducible quantitative experiments require logs that can be analyzed by scripts. QUANTAS~2 records structured JSON output for each experiment and test. The recorded parameters include: seeds, host metadata, round-level metrics, and algorithm-specific values. Additionally, a category-based logger provides error, warning, info, debug, and trace levels so that the same run can produce compact experimental data and detailed debugging traces.

\ \\
\textbf{Scalability.}
Once the basic behavior of a distributed algorithm is ascertained, researchers usually want to observe its behavior at scale. To  support this, QUANTAS~2 allows peers to be distributed across several computers in a concrete simulation. Potentially, increasing the simulated network size as it is not limited by the  processor or memory resources of a single host computer.

\begin{figure*}
   \centering
    \includegraphics[width=0.9\textwidth]{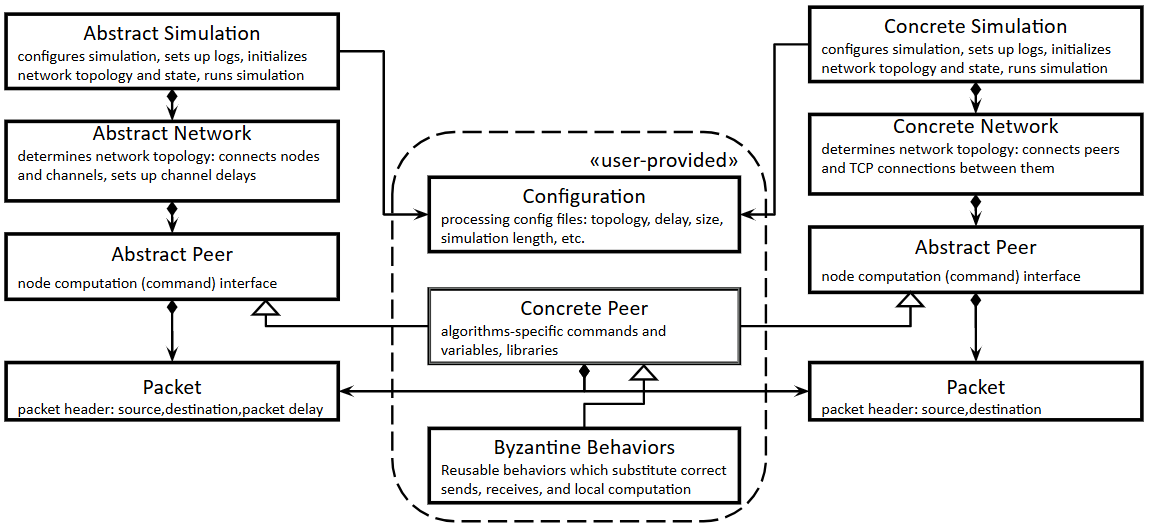}
    \caption{QUANTAS~2 architecture.}
    \label{figArch}
\end{figure*}

\ \\
\textbf{Terms and operation.}
A simulated \emph{distributed algorithm} operates on a list of nodes connected via unicast channels. Each channel connects a single sender and a single receiver.  Every node has a unique identifier. 
Each  \emph{computation} of the distributed algorithm is a sequence of \emph{rounds}. Each round has two \emph{phases}: receive messages, and perform local computation. A computation \emph{length} is its number of rounds.
A message takes at least one round to pass between nodes that are directly connected through a channel. A message can be delayed. Delay length is configured. The delay is also configured to be either deterministic, uniformly random, or following a Poisson distribution. Communication channels are FIFO by default. Other message propagation delay disciplines may be added by the user.
A transmitted message may be configured to be lost with a certain probability. A message may be sent to an individual node or broadcast to the entire network.
A single \emph{run} of the QUANTAS~2 simulator executes several algorithm computations with the same parameters. This allows QUANTAS~2 to execute multiple individual experiments for a single data point. 

\subsection{Architecture}
QUANTAS~2 architecture is shown in Figure~\ref{figArch}. The components represent the larger C++ templates and classes. 
The components are in two categories: user-provided and the simulator proper. The user-provided components encode the algorithm to be simulated. The simulator components carry out the simulation. The run-time operation of the simulator is controlled by configuration files.

The  \emph{Simulation Component} configures and initializes the simulation run. It then carries out the receive-compute-send computation rounds of individual computations of the run. The Simulation Component uses the \emph{Configuration Component} for processing user-supplied configuration file containing network topology and size, parameters of the run, message delay discipline and parameters, computation length, etc. The network topology is specified as adjacency list and can be generated by hand or by a separate tool. 

Since the execution of the same round in the separate simulated nodes is not causally related,
this execution is done concurrently by separate threads. To carry this out, the Simulation Component maintains a thread pool. By default, the number of threads in this thread pool is the maximum  that the host computer can concurrently run. 

The \emph{Network Component} configures distributed algorithm topology, sets up communication channels and executes receive- compute- and send- phases of the round. The \emph{Abstract Node Component} is a C++ abstract class that lists the interfaces to be implemented by a user-provided \emph{Concrete Node} component. The main part of this interface is the code to be executed in local computation phase of the round.  

The \emph{Node Network Interface Component} executes receive and send phases of the round. In the receive phase, The Node Network Interface Component examines all the channels, and determines if any of the messages currently in transit are ready to be received. The ready-to-receive messages are made available for the computation phase. If the computation phase generates messages to be transmitted, the Node Network Interface Component collects them and puts them in the appropriate destination channels.

A message is enclosed in a packet. The packet contains the source, destination, and the delay for this particular message. The \emph{Packet Component} provides this header and the Node Network Interface Component uses this header for message routing. The actual message format and its payload are provided by the \emph{User-Defined Message Component}. 

Let us discuss QUANTAS~2 data output capabilities. QUANTAS~2 provides a global logging facility, so that each component may output to the log file. All simulator components may output data about their particular operation. For example, the Node Network Interface Component may output sender and receiver identifiers for each individual message. The user-provided components may output arbitrary data, which enables user-specific metrics to be easily implemented. User-provided components have access to the computation round number maintained by the simulator. This round number can be included in the output for analysis. To simplify later processing, logger allows to attach an arbitrary tag to output lines.

\subsection{Concrete Mode}
\begin{comment}
% at this point int he paper, the reader is already sold on need for concrete simulation, I do not think we need this here, MN
The original QUANTAS execution model is abstract: nodes exchange packets through simulator-managed channels and progress in synchronous rounds. This is the appropriate default for algorithmic studies because it provides deterministic control over topology, delay distributions, queue sizes, loss, duplication, and reordering. However, once an algorithm has been implemented, researchers often want to know whether the same code can run as a networked process without rewriting the experiment.
\end{comment}
Concrete mode allows to execute the algorithm code as a networked process. In concrete mode, QUANTAS~2 uses socket-based network interfaces and a process coordinator rather than purely simulated channels. A process may act in one of the two \emph{roles}. A single \emph{leader process} listens on a configured port. \emph{Follower processes} connect to the leader and participate in the same peer-level algorithm. The concrete mode can be launched on a single computer for debugging purposes or across a list of pre-configured remote computers.
%through the provided makefile routes. 
QUANTAS~2 records the computer name, computer IP, process role, and output directory.  This makes distributed experiment output easier to associate with the process that produced it.

% move this to the intro? MN
% Mikhail Stopped here

\subsection{Byzantine Fault Simulation} 

Potentially, Byzantine fault behavior is quite complex. A Byzantine node may send conflicting values to different quorums, withhold blocks until a strategic release point, rewrite selected fields, ignore received messages, or replace its local computation entirely. 
QUANTAS~2 allows researchers to implement arbitrary Byzantine faults by overriding correct node operation. Specifically, 
a \emph{fault object} can override four \emph{hooks}: direct unicast, multicast/broadcast send, receive, and local computation. 
%The fault manager attached to a peer dispatches these hooks before the ordinary peer behavior.
If a hook from the fault object returns that it handled the action, the correct process action is suppressed
%; otherwise the algorithm proceeds normally.

\begin{figure}
    \centering
    \begin{lstlisting}
class EquivocateFault : public Fault {
 bool onSend(Peer* peer,
                json& msg,
                const std::set<interfaceId>& 
                targets = {}) override
    {
    // Find the midpoint
    auto mid = std::next(targets.begin(), 
        targets.size() / 2);
    
    // First Half: [begin, mid)
    std::set<int> firstHalf(targets.begin(), mid);
    
    // Second Half: [mid, end)
    std::set<int> secondHalf(mid, targets.end());

    // Send differing messages
    peer->getNetworkInterface()->multicast(msg,
        firstHalf);
    json alt = msg; alt.var = !msg.var;
    peer->getNetworkInterface()->multicast(alt,
        secondHalf);
};
\end{lstlisting}
    \caption{Example Byzantine behavior of an equivocation fault used in PBFT to substitute a multicast. Sends the intended message to the first half of the targets and a message with the opposite value to the second half.}
    \label{figFaultHooks}
\end{figure}

Presently, QUANTAS~2 implements two reusable Byzantine behaviors. The \emph{equivocation fault} partitions a multicast target set into two quorums and sends conflicting payloads to each side. This is especially useful for PBFT-style algorithms, where safety depends on quorum intersection and equivocation can cause replicas to observe incompatible certificates~\cite{PBFT,lamport2019byzantine}. 

The \emph{parasite fault} models a selfish proof-of-work coalition that may be perpetrated in blockchains. Adversarial miners withhold mined blocks, exchange them privately with collaborators, continue mining on the private branch, and release the hidden chain once it reaches a configured lead. This behavior is motivated by selfish mining and topology-sensitive majority attacks, where network propagation and peer placement can let an attacker with less than half of the mining power gain disproportionate influence~\cite{eyal2014majority,gervais2016security,saad2021revisiting}.

QUANTAS~2 Byzantine fault simulation is similar in spirit to recent BFT testing systems that generate Byzantine scenarios without requiring the algorithm implementation itself to be malicious. Twins, for example, emulates Byzantine behavior by running multiple instances with the same identity and credentials~\cite{bano2020twins}. QUANTAS~2 instead exposes explicit programmable hooks inside a simulation and concrete-execution harness. The two approaches are complementary: Twins is attractive for testing production BFT implementations with minimal code changes, while QUANTAS~2 is designed for parameter sweeps, topology studies, and repeatable quantitative experiments during algorithm development. 
% Joseph, fill!
Our experimental data is available to download~\cite{}.

\section{Simulation Examples}

In this section, we demonstrate how QUANTAS~2 may be adapted to fit diverse experimentation needs of distributed algorithms researchers. 
We chose four domains and a pair of previously published well-known algorithms in each domain. We then implemented the algorithms in QUANTAS~2 and compared their performance. The first three pairs of algorithms are implemented in the abstract mode. We injected faults and observed how the algorithms handle them. The last pair of algorithms is implemented in both the abstract and the concrete mode. We observed the speedup achieved by the concrete mode implementation. While the results themselves are not surprising, they demonstrates QUANTAS~2 simulation capabilities on the variety of distributed algorithms. 

%We also evaluated the speed of QUANTAS~2 simulation execution. Specifically, we measured the speedup achieved by QUANTAS~2 as threads are added to the concurrent thread pool allowing greater concurrency and faster execution. 
%

\begin{figure}
   \centering
   \includegraphics[width=0.49\textwidth]{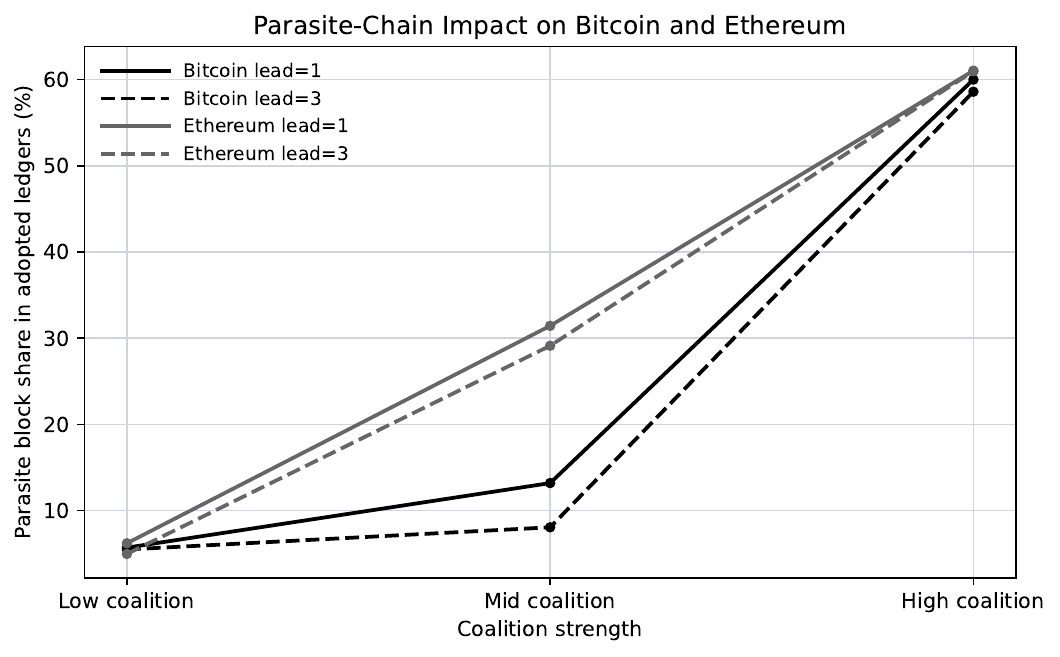}
   \caption{Parasite-chain Byzantine attack for Bitcoin and Ethereum algorithm.}
   \label{figParasiteSweep}
\end{figure}

\subsection{Blockchains}

Blockchain is a secure distributed ledger maintained by a network of peers that compete to add blocks of transactions to the tail of the chain. We simulate simplified versions of Bitcoin~\cite{bitcoin} and Ethereum~\cite{ethereum} on a 12-peer complete network. Each peer maintains its own copy of the ledger. Transactions are submitted continuously through broadcast, and miners extend the ledger according to the local protocol rules. In our implementation, a Bitcoin miner extends the longest block chain known to it. An Ethereum miner uses a GHOST-style subtree-weight rule to choose the preferred tip to which it adds a newly mined block~\cite{sompolinsky2021phantom}.

We evaluate these algorithms under a parasite-chain fault. The \emph{coalition} of two colluding miners withhold newly mined blocks from honest peers, exchange those blocks privately with one another, and continue mining on the hidden branch. The hidden branch is not released immediately. Instead, we use a two-part release rule. First, the honest public chain must advance by at least a configured public-progress threshold beyond the private branch's starting base. In our experiments this parameter is fixed at $\texttt{publicBlockThreshold}=2$. Second, the hidden branch must be ahead of the current public chain by at least a configured lead threshold. We use $\texttt{leadThreshold}\in\{1,3\}$. 

We simulate three \emph{coalition strength} by varying the mining rates of the two parasite peers relative to the ten honest peers. In the low strength coalition, the two attackers have a mining rate of $2$ while each honest peer has a rate of $1$. That is, a attacker is twice as likely to mine a block as an honest miner. In the medium strength coalition, the attackers have a mining rate of $4$, In the high strength, they have $6$. 

Figure~\ref{figParasiteSweep} shows the fraction of parasite blocks that appear in the ledgers ultimately adopted by the peers.
The measured behavior varies sharply with attackers' strength. At low coalition strength, both Bitcoin and Ethereum remain near $5\%$ to $6\%$ parasite share. At high coalition strength, both exceed $58\%$, showing that the private branch is frequently able to dominate the public branch. The most interesting separation occurs in the middle strength coalition: Bitcoin remains near $8\%$ to $13\%$, while Ethereum rises to roughly $29\%$ to $31\%$. This difference comes from the fact that the  simulator's Ethereum variant does not use Bitcoin's longest-chain rule. Instead, it uses GHOST. As a result, a parasite branch can become preferred in Ethereum once it contributes enough descendant weight to the favored subtree, even though honest miners still extend only a single selected parent at each step. This makes it so that the attackers coalition strength easier to combine and overtake the honest miners.

The release thresholds also behave consistently with the implemented fault. A lead threshold of $1$ is more aggressive than a lead threshold of $3$: once the public chain has advanced by the required two blocks, the coalition reveals as soon as its private branch is only one block ahead, rather than waiting for a larger safety margin. This earlier release generally produces a stronger attack in the medium coalition because the coalition loses fewer races to additional honest-public growth, whereas a lead threshold of $3$ waits longer and therefore misses more opportunities to replace the public branch.

These experiments are motivated by the gap between the idealized ``51\% attack'' rule of thumb and the behavior of real proof-of-work networks. 
That is, it is often assumed that blockchains are resilient to adversarial attack as long as the honest miners posess at least $51\%$ of hash power. However, selfish mining showed that a miner with less than half of the hash power can obtain more than its proportional revenue by strategically withholding blocks~\cite{eyal2014majority}. Subsequent work showed that propagation delay, block size, and peer connectivity affect the security margin of Nakamoto consensus~\cite{decker2013information,gervais2016security}. More recent topology-sensitive analyses show that adversarial placement and asymmetric delays can further reduce the effective cost of majority-style attacks~\cite{saad2021revisiting}. Our experiments demonstrate that QUANTAS~2 is well suited for exploring these questions because QUANTAS~2 can vary mining power, delay distribution, and network graph independently while keeping the implementation fixed.

\begin{figure}
\centering
\includegraphics[width=0.49\textwidth]{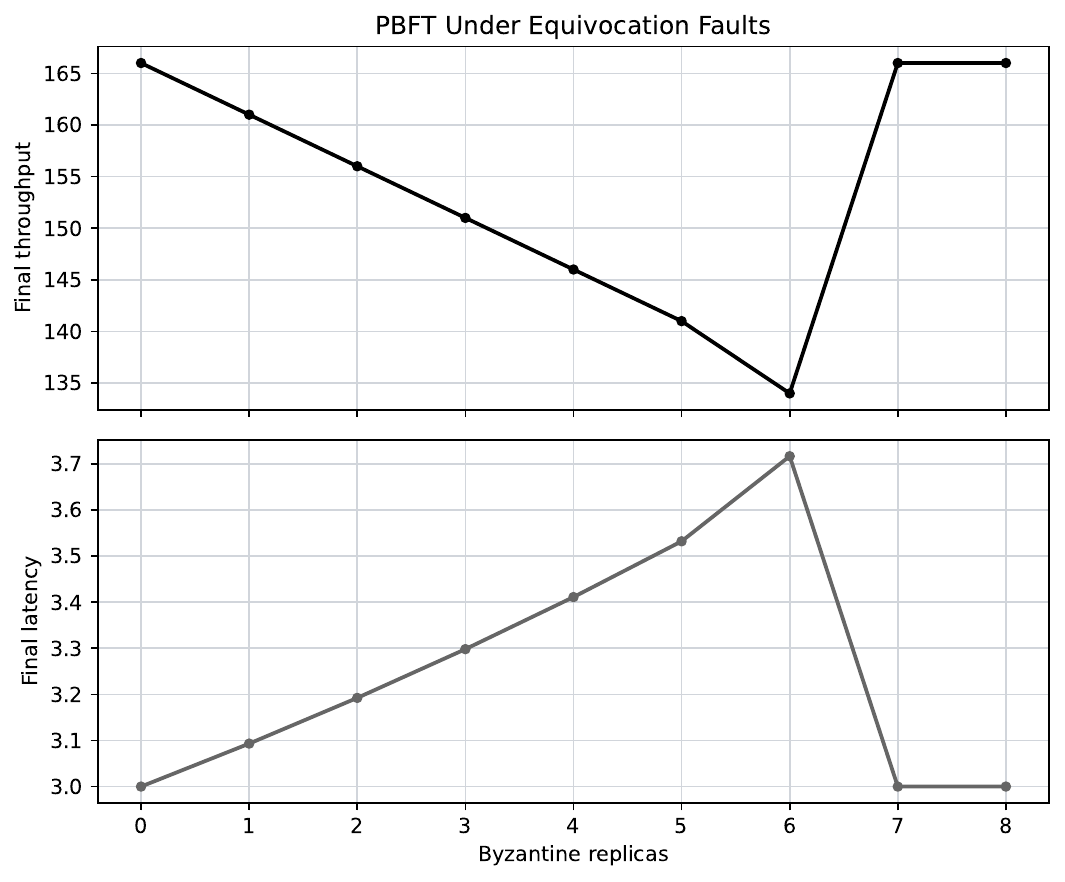}
\caption{PBFT under equivocation faults.}
\label{figPBFTByzantine}
\end{figure}

\begin{figure}
\centering
\includegraphics[width=0.49\textwidth]{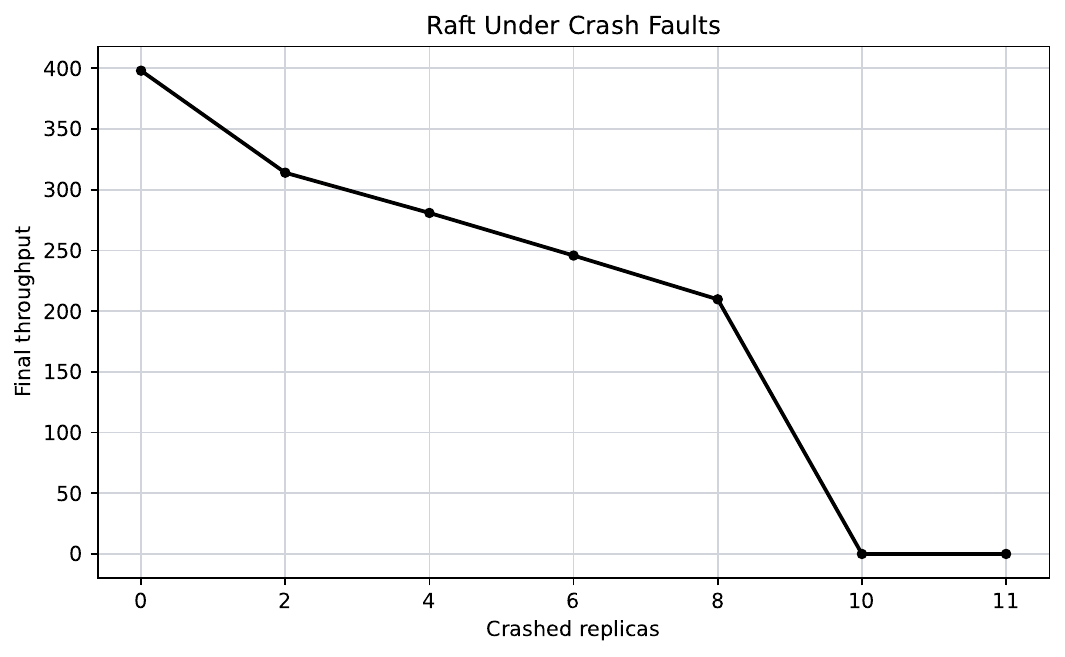}
\caption{Raft under crash faults and recovery.}
\label{figRaftCrash}
\end{figure}

\begin{figure}
\includegraphics[width=0.49\textwidth]{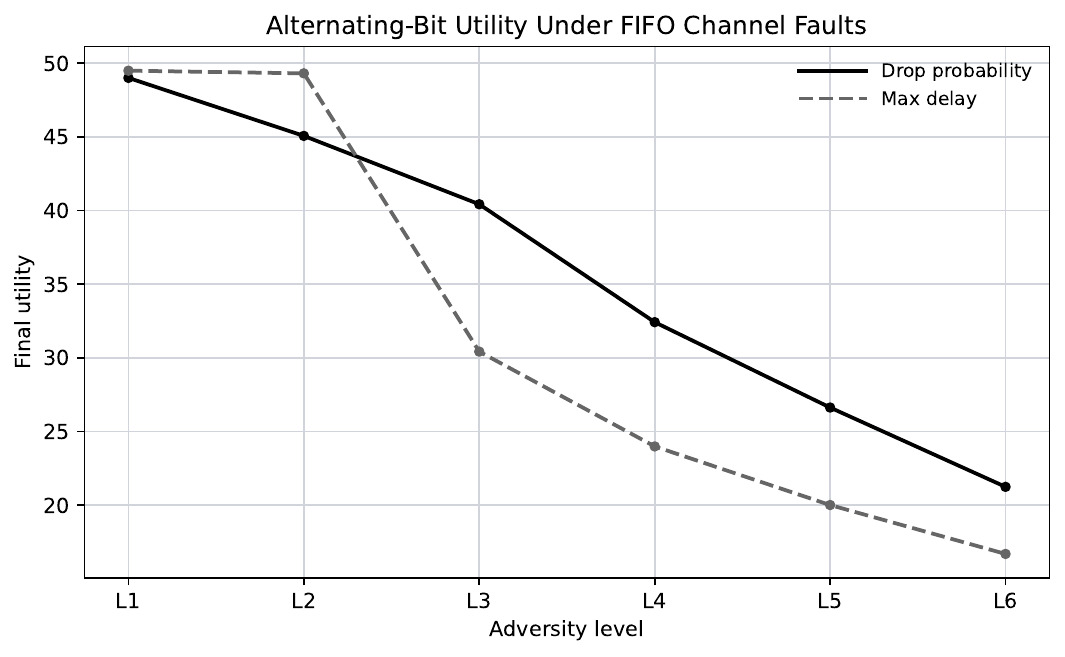}
\caption{Alternating-bit utility under FIFO channel faults.}
\label{figAltBit}
\end{figure}

\begin{figure}
\includegraphics[width=0.49\textwidth]{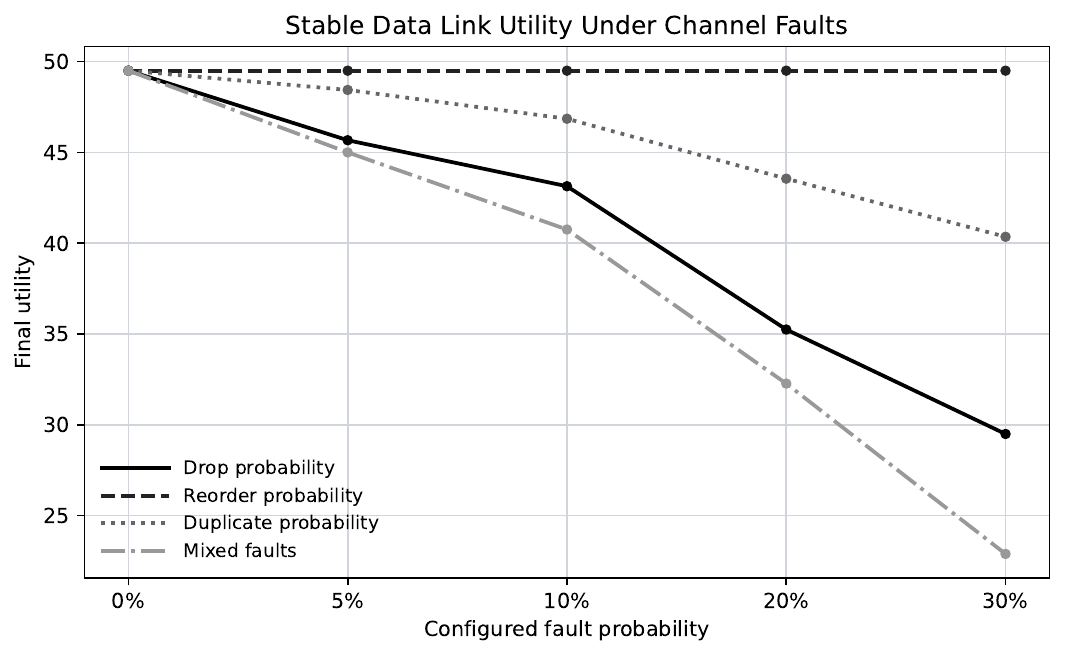}
\caption{Stabilizing data link utility under channel faults.}
\label{figSDL}
\end{figure}

\subsection{Robust Consensus} 

\textbf{Equivocation faults.}
In robust consensus, a network of nodes attempts to agree on a sequence of requests despite failures. PBFT~\cite{PBFT} and Raft~\cite{raft} are two of the most popular consensus robust consensus algorithms. PBFT is resilient to Byzantine faults. There is a single primary node that orders the requests while the rest of the nodes, called replicas, confirm that order by comitting the requests. In the primary node fails, then a new primary is selected. In PBFT, the process is called a view change. 

In our implementation, the network begins with a designated primary $l$, but may rotate to later primaries through PBFT view changes when timeouts occur. For each value, the primary broadcasts a \emph{pre-prepare} message, replicas answer with \emph{prepare}, and then \emph{commit} messages finalize the decision. For our experiments, we used $19$ nodes and executed each computation for $500$ rounds with $10$ tests per data point. 

The PBFT equivocation fault experiment attaches equivocation faults to a configurable number of replicas. A faulty replica sends the normal value to one quorum and a conflicting value to another quorum. The experiment records final throughput and latency. The results are shown in Figure~\ref{figPBFTByzantine}. The data indicates that PBFT behaves as expected through the algorithm's operating range: the throughput declines gradually from $166$ committed values at $0$ Byzantine replicas to $134$ at $6$ Byzantine replicas. The throughput rises sharply after the number of faulty replica increase past $7$ since this number is greater than PBFT fault tolerance threshold and the algorithm is defeated. The latency declines accordingly.

\ \\
\textbf{Crash faults.} In Raft, the leader node broadcasts requests to all nodes and moves to the next value after receiving a majority of responses. If a leader node crashes or takes too long to respond to the other nodes then a new leader is elected. Raft is designed for crash fault tolerance only.

Our Raft experiment evaluates crash node.  The experiment varies the number of crashed replicas and records final throughput. The results are shown in Figure~\ref{figRaftCrash}. We simulate $21$ replicas for $800$ rounds with $10$ tests per data point. The data shows a smooth throughput  decline  as the number of crashed replicas increases from $0$ to $8$, followed by complete loss of progress once the crash count reaches $10$ and the system can no longer maintain the majority needed for consensus. The contrast between Figures~\ref{figPBFTByzantine} and~\ref{figRaftCrash} is instructive: PBFT pays additional communication cost to tolerate equivocation, while Raft is faster in benign executions but assumes faulty nodes stop rather than lie.

\subsection{Reliable Data Link}

\textbf{Message drop.}
In a data link algorithm, the sender node attempts to transmit data to the receiver node despite adverse channel behavior. A self-stabilizing algorithm~\cite{dijkstra1974self} is resilient to global state corruption. We implemented two fundamnetal self-stabilizing data link algorithms: alternating-bit protocol (ABP)~\cite{howell2002finite} and stabilizing-data link algorithm (SDL)~\cite{dolev2011stabilizing}. 

ABP requires FIFO channels. In ABP, the sender transmits a single data message and waits for acknowledgment from the receiver. If either the data message or the acknowledgment is dropped (lost), the sender times out and retransmits the message.

In our simulation, we used $2$ nodes, ran each computation for $200$ rounds, and executed $10$ tests per data point. We report \emph{message utility}: successfully delivered messages divided by transmitted messages, expressed as a percentage. 
For ABP, we varied drop probability and maximum delay. The timeout is fixed at $3$ rounds. 

See Figure~\ref{figAltBit} for ABP results. ABP responds similarly to message delay and message drop. Threfore, we combine our results for message delay and message drop probability into 6 adversity levels: $L1$ through $L6$ and measure utility level for both. The relationship between adversity levels and specific drop probabilities and message delays is shown in Table~\ref{levelTable}. 

The points for delay levels $L1$ and $L2$ are very close, while the curve bends sharply downward at $L3$. The data confirms the effect of message drop and message delay on APB is similar.

\begin{table}%[htp]
\centering
\begin{tabular}{|c|cccccc|}
\hline
adversity level & 1 & 2 & 3 & 4 & 5 & 6 \\
\hline
delay & 1 & 2 & 3 & 4 & 5 & 6 \\
drop probability & 0 & 0.05 & 0.1 & 0.2 & 0.3 & 0.4 \\
\hline
\end{tabular}
\caption{Adversity levels with corresponding delay and drop probability.}
\label{levelTable}
\end{table}

\begin{comment}
This is because in our experiment, $L1$ has a one-way delay of $1$ and $L2$ has a one-way delay uniformly distributed in $\{1,2\}$, with timeout rate $3$. Because the round-trip time at $L2$ is usually still below the timeout threshold, acknowledgments typically arrive before retransmission triggers. As a result, throughput drops from $99.0$ to $64.8$ completed messages, but utility remains almost unchanged: the sender is waiting longer, not wasting many more messages. At $L3$, the round-trip delay can exceed the timeout, so retransmissions become common and utility falls to about $30.4\%$. The drop-probability line behaves differently: packet loss directly forces retransmission even when delay is small, so utility declines more smoothly as drop probability increases.
\end{comment}

\ \\
\textbf{Combined message drop, reorder and duplication.}  SDL operates correctly even in non-FIFO channels. It uses message sequence numbers to restore message order. 

%In SDL, to enforce ordered delivery under weaker channel assumptions, the sender intentionally sends additional control traffic and redundant messages.

To evaluate SDL, we retained the experimental parameters of ABP and varied the probability of drop, reorder, duplicate. Mixed channel faults is the combination of all fault types. Similar to ABP, SDL sends only one message at a time. The results of our SDL experiments are shown in Figure~\ref{figSDL}.

Figure~\ref{figSDL} shows that SDL reacts differently to different non-FIFO fault types. Message reordering has no observable effect on SDL as the sender sends only one message at a time and retransmits the same message after the timeout, so out-of-order delivery does not change which message can be received next. 
Duplicate faults moderately reduce utility because the algorithm still delivers data correctly while using extra bandwidth for retransmission. 
Drop faults are more harmful, and the mixed-fault combination is the worst overall because it combines loss with duplication and creates the potential to reorder messages in the event of duplications.
Figure~\ref{figSDL} highlights QUANTAS~2 capabilities of evaluating the same algorithm under different faulty types. 

\subsection{Distributed Hash Tables} 

\begin{figure}
\centering
\includegraphics[width=0.49\textwidth]{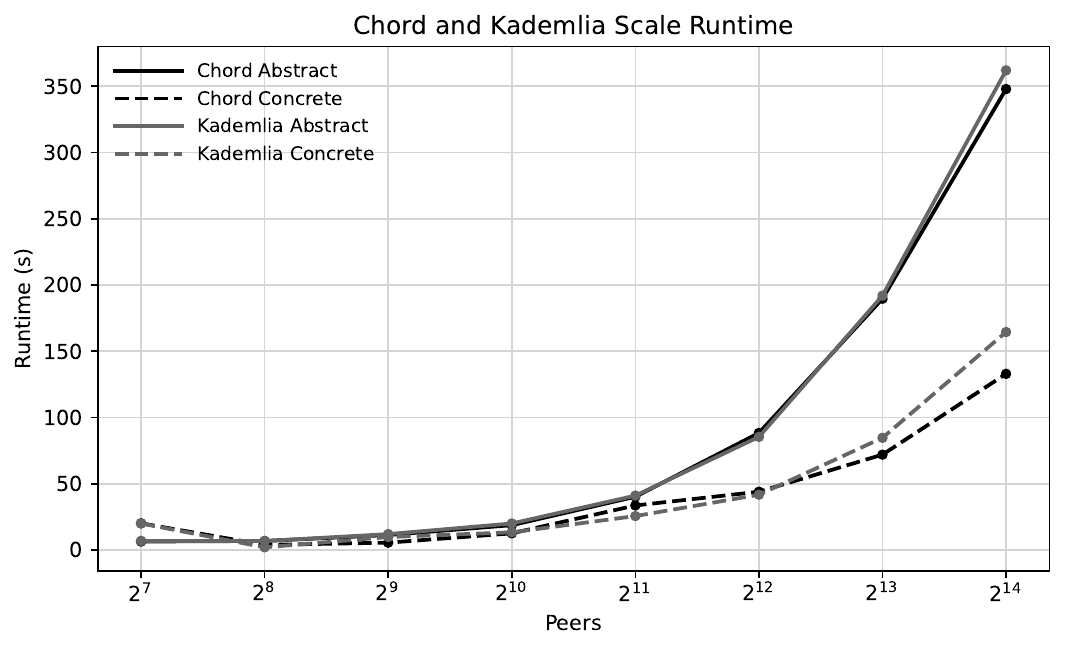}
\caption{Chord and Kademlia scale runtime in abstract and concrete modes.}
\label{figSparseScale}
\end{figure}

\textbf{Concrete and abstract simulation at scale.} In a distributed hash table (DHT), a peer-to-peer system provides query service for key to data items spread throughout the network. The algorithm is optimized to minimize the number of lookups per query. Some of the most widely used DHTs are Chord~\cite{chord} and Kademlia~\cite{kademlia}.

In our Chord implementation, the peer identifiers form a ring. A query for an identifier chosen uniformly at random is generated by another random node. The query is routed to the destination node in the shortest direction using short cut links. 

In Kademlia we build our links as follows. Peer identifiers are treated as bit fields. A prefix peer group for a particular peer $p$ is a set of peers whose identifiers share a prefix of particular length $l$ with $p$ and differ from $p$ at length $l+1$. For example, if the prefix is one bit less than the complete id length, then, there is a single member in this peer group. A peer group for a prefix that is two bits shorter than id length, contains two members. A peer group three bit shorter than id length contains 4 members and so on.  For each group, a peer selects a random member and creates a shortcut link to it. The query routing is as follows. The peer selects a member with the closest prefix to the destination and routes the query there. 

We used the same implementation for abstract and concrete mode experiments. For the concrete experiments, we used $10$ virtual machines (VMs) each with $10$ CPU cores running Intel Xeon Gold series CPUs $6340$ @ $2.80$ GHz, $6240$ @ $2.6$GHz and $6242$ @ $3.1$ GHz and $64$GB of RAM. The VMs were connected by $20$ GB Ethernet Aruba 8325 network switch. 

We varied the number of DHT peers from $2^7$ to $2^{14}$ . In each test, we carry out $100$ queries. For both the abstract and concrete implementations, we measured the physical runtime. We run $10$ tests for each data point and sum the elapsed time. The results are shown in Figure~\ref{figSparseScale}.

All four series rise with scale, however, concrete mode simulations are significantly faster.  In abstract mode, runtime rises from roughly $4$ to $5$ seconds at $128$ peers to about $347$ seconds for Chord and $362$ seconds for Kademlia at $16{,}384$ peers. The experiments demonstrate 
QUANTAS~2 capability to handle large scale simulated networks as well as the benefits of using multiple computers run simulations.

\section{Conclusions and Future Work}
QUANTAS~2, presented in this paper, gives distributed algorithm researchers a single platform in which they can move from controlled abstract studies to concrete executions and adversarial experiments while reusing the algorithim and experiment structure. By reducing the gap between algorithm design, performance evaluation, and fault-oriented testing, QUANTAS~2 helps make quantitative distributed-systems research more reproducible, comparable, and extensible.

\ \\
Over the four years since the introduction of the original QUANTAS, the simulator has proven quite useful. QUANTAS~2, presented in this paper, 
takes it to the next level of usability. However, we do not believe we exhausted the opportunities for its further enhancement. 

The present version supports message loss, crash-style behavior, and programmable Byzantine fault strategies. A natural next step is a richer library of reusable faults: omission faults, timing faults, coordinated partitions, and adaptive adversaries. 

%In particular, we plan to add support for self-stabilizing algorithms evaluation. Even though a self-stabilizing algorithm is proven to recover from an arbitrary global state, evaluating the algorithm's performance starting from a state generated uniformly at random is not realistic as not all such states are equally likely to appear. A more sophisticated approach was developed by Adamek et al~\cite{adamek12sss}: an achievable state of a self-stabilizing algorithm is selectively perturbed. We would like to implement this kind of fault injection in QUANTAS~2.

The Byzantine interface also opens the possibility of adversary search. At present, the researcher specifies a fault strategy and sweeps its parameters. Future work could combine QUANTAS~2 with randomized testing, model checking, or reinforcement learning to discover schedules and faulty actions that maximize latency, safety violations, or wasted work. This would make QUANTAS~2 useful not only as a measurement harness but also as a tool for finding weaknesses before deployment.

As a further enhancement, we would like to add the ability to change the network parameters during a single computation. This would allow researchers to model mobile or dynamic networks where the message delay might vary throughout the experiment. 

Another feature we find useful is facilitation of application level separation. This would allow the simulator to evaluate levels of multi-level algorithms separately, for example, evaluate the same consensus algorithm over different broadcast algorithms.

% In this paper, we presented QUANTAS~2, a general abstract and concrete simulator dedicated to distributed algorithms quantitative evaluation. 
% While we provided a number of case studies, we welcome contributions from the Distributed Computing community, to build a library of ready-to-use templates for most algorithmic paradigms, that enables fair comparison with previous work when designing new solutions. We believe that QUANTAS~2 fulfills the need for a simulator among researchers of distributed algorithms and we hope it proves to be useful and convenient. 

\bibliographystyle{plain}
\bibliography{quantas}

\end{document}